\documentclass[%
reprint,
superscriptaddress,
footinbib,
twocolumn,
amsmath,amssymb,
aps,
prl,
floatfix,
longbibliography
]{revtex4-1}

\usepackage{amsmath}
\usepackage{graphicx}
\usepackage{dcolumn}
\usepackage{bm}
\usepackage{ulem}
\usepackage{amssymb}
\usepackage{multirow}
\usepackage[colorlinks=true,citecolor=blue]{hyperref}
\begin{document}

\title{Possible Kitaev Quantum Spin Liquid State in 2D Materials with $S=3/2$}

\author{Changsong Xu}
\altaffiliation{Contributed equally to this work.}
\affiliation{Physics Department and Institute for Nanoscience and Engineering, University of Arkansas, Fayetteville, Arkansas 72701, USA}%

\author{Junsheng Feng}
\altaffiliation{Contributed equally to this work.}
\affiliation{Key Laboratory of Computational Physical Sciences (Ministry of Education), State Key Laboratory of Surface Physics, and Department of Physics, Fudan University, Shanghai, 200433, China}%
\affiliation{School of Physics and Materials Engineering, Hefei Normal University, Hefei 230601, P. R. China}%

\author{Mitsuaki Kawamura}
\affiliation{The Institute for Solid State Physics, The University of Tokyo, Kashiwa-shi, Chiba, 277-8581, Japan}

\author{Youhei Yamaji}
\affiliation{Department of Applied Physics, University of Tokyo, Hongo, Bunkyo-ku, Tokyo 113-8656, Japan}

\author{Yousra Nahas}
\affiliation{Physics Department and Institute for Nanoscience and Engineering, University of Arkansas, Fayetteville, Arkansas 72701, USA}%

\author{Sergei Prokhorenko}
\affiliation{Physics Department and Institute for Nanoscience and Engineering, University of Arkansas, Fayetteville, Arkansas 72701, USA}%

\author{Yang Qi}
\email{qiyang@fudan.edu.cn}
\affiliation{Key Laboratory of Computational Physical Sciences (Ministry of Education), State Key Laboratory of Surface Physics, and Department of Physics, Fudan University, Shanghai, 200433, China}%

\author{Hongjun Xiang}
\email{hxiang@fudan.edu.cn}
\affiliation{Key Laboratory of Computational Physical Sciences (Ministry of Education), State Key Laboratory of Surface Physics, and Department of Physics, Fudan University, Shanghai, 200433, China}%
\affiliation{Collaborative Innovation Center of Advanced Microstructures, Nanjing 210093, P. R. China}

\author{L. Bellaiche}
\affiliation{Physics Department and Institute for Nanoscience and Engineering, University of Arkansas, Fayetteville, Arkansas 72701, USA}%


\begin{abstract}
  Quantum spin liquids (QSLs) form an extremely unusual magnetic state in which the spins are highly correlated and fluctuate coherently down to the lowest temperatures, but without symmetry breaking and without the formation of any static long-range-ordered magnetism. Such intriguing phenomena are not only of great fundamental relevance in themselves, but also hold the promise for quantum computing and quantum information. Among different types of QSLs, the exactly solvable Kitaev model is attracting much attention, with most proposed candidate materials, e.g., RuCl$_3$ and Na$_2$IrO$_3$, having an effective $S$=1/2 spin value.
  Here, via extensive first-principle-based simulations, we report the investigation of the Kitaev physics and possible Kitaev QSL state in epitaxially strained Cr-based monolayers, such as CrSiTe$_3$, that rather possess a $S$=3/2 spin value.
  Our study thus extends the playground of Kitaev physics and QSLs to 3$d$ transition metal compounds.
 \end{abstract}
\maketitle

Enormous efforts have been made to realize quantum spin liquids (QSLs) since the pioneering work of Anderson and Fazekas in 1970s\cite{anderson1973resonating,fazekas1974ground}. Models with typical ingredients, such as geometrical frustration and antiferromagnetism (AFM), have been extensively studied\cite{balents2010spin}.
Recently, the two-dimensional (2D) Kitaev model defined on a honeycomb lattice has attracted much attention, since its ground state is exactly proved to be QSL with Majorana fermion excitations\cite{kitaev2006anyons}. Later work by Jackeli and Khaliullin demonstrated that such model can be realized in certain transition metal compounds with strong spin-orbit coupling (SOC)\cite{jackeli2009mott}. Measurements also observed proximate Kiteav QSL in candidate materials, such as  $\alpha$-RuCl$_3$\cite{banerjee2016proximate,baek2017evidence,zheng2017gapless} and (Na$_{1-x}$Li$_x$)$_2$IrO$_3$\cite{cao2013evolution,manni2014effect}, which all have a spin state of effective $S$ = 1/2, as well as, a layered structure with edge-sharing octahedras and strong SOC from 4$d$ or 5$d$ transition metals.
However, to the best of our knowledge, unambiguous demonstration of the Kitaev QSL is still lacking and zigzag ordered state tends to form in  the aforementioned materials at very low temperatures.
Possible reasons are that (i) the structure distorts away from ideal honeycomb lattice and (ii) the relative strengths of isotropic exchange coupling and Kitaev interaction plays a role in the forming of ground states\cite{chaloupka2013zigzag,yamaji2016clues}.
It is thus necessary to search for other candidates and/or find a way to tune magnetic interactions, in order to create QSL systems.

Recently,  atomic layers made of CrI$_3$ and CrGeTe$_3$ have been  synthesized and found to be ferromagnetic, which  resulted in a major surge of researches dedicated to two-dimensional (2D) magnetism\cite{huang2017layer,gong2017discovery}. CrI$_3$ and CrGeTe$_3$  share similarities in their crystal structure with $\alpha$-RuCl$_3$ and Na$_2$IrO$_3$, i.e., they all have a honeycomb lattice and edging-sharing octahedra.  On the other hand, CrI$_3$ and CrGeTe$_3$ have a higher spin state than $\alpha$-RuCl$_3$ and Na$_2$IrO$_3$, namely $S$ = 3/2 {\it versus} $S$ = 1/2. At first thought, the FM nature and large $S$ value, as well as the light SOC associated with Cr, seemingly exclude CrI$_3$ and CrGeTe$_3$ from being Kitaev QSL candidates.
However, one of our recent works\cite{xu2018interplay} hints that these systems exhibit finite Kitaev interaction that arises from heavy ligands of I/Te and thus may in fact be promising to find QSL.
As a matter of fact, it is important to know that (1) such previous work adopted
 the general matrix form of Hamiltonian
\begin{equation}
 \mathcal{H} = \frac{1}{2} \sum_{i,j} \bm{{\rm S}}_i {\cdot} \mathcal{J}_{ij} {\cdot} \bm{{\rm S}}_j + \sum_{i} \bm{{\rm S}}_i {\cdot} \mathcal{A}_{ii} {\cdot} \bm{{\rm S}}_i
\end{equation}
where the first sum runs over all nearest neighbors and the second sum runs over all single sites; and (2) the $\mathcal{J}_{\rm X}$, $\mathcal{J}_{\rm Y}$ and $\mathcal{J}_{\rm Z}$ matrices respectively have the following forms\\
$
 \left({\begin{array}{*{20}{c}}
J$+$K&\mathit{\Gamma}_2&\mathit{\Gamma}_2\\
\mathit{\Gamma}_2&J&\mathit{\Gamma}_1\\
\mathit{\Gamma}_2&\mathit{\Gamma}_1&J\\
\end{array}}\right)
\left({\begin{array}{*{20}{c}}
J&\mathit{\Gamma}_2&\mathit{\Gamma}_1\\
\mathit{\Gamma}_2&J$+$K&\mathit{\Gamma}_2\\
\mathit{\Gamma}_1&\mathit{\Gamma}_2&J\\
\end{array}}\right)
\left( {\begin{array}{*{20}{c}}
J&\mathit{\Gamma}_1& \mathit{\Gamma}_2\\
\mathit{\Gamma}_1&J&\mathit{\Gamma}_2\\
\mathit{\Gamma}_2&\mathit{\Gamma}_2&J$+$K\\
\end{array}}\right)$\\
implying  that Eq. (1) can be rewritten as:
\begin{equation}
\begin{aligned}
\mathcal{H}&=\frac{1}{2}\sum_{i,j}\{J\bm{{\rm S}}_i{\cdot}\bm{{\rm S}}_j + KS_i^{\gamma}S_j^{\gamma} +
\mathit{\Gamma}_1(S_i^{\alpha}S_j^{\beta} + S_i^{\beta}S_j^{\alpha})+\\
&\mathit{\Gamma}_2(S_i^{\gamma}S_j^{\alpha} + S_i^{\gamma}S_j^{\beta}+
S_i^{\alpha}S_j^{\gamma} + S_i^{\beta}S_j^{\gamma})\}+
\sum_{i} A_{zz}{\rm S}^z_i{\rm S}^z_i
\end{aligned}
\end{equation}
where $\{\alpha,\beta,\gamma\}$=$\{Y,Z,X\}$, $\{Z,X,Y\}$ and $\{X,Y,Z\}$ for the X-, Y- and Z-bonds, respectively. Note that
the global $\{XYZ\}$ basis and the X-, Y- and Z-bonds are shown in Figure 1, and that only the $A_{zz}$ component is finite in the $\mathcal{A}$ matrix when expressed in the global $\{xyz\}$ basis -- which explains why only $A_{zz}$ appears in the SIA term. Interestingly,
Eq. (2) characterizes a typical $JK\mathit{\Gamma}$ model\cite{winter2017models} but with an with extra SIA term. Consequently, Eq. (2) represents what we coin here as a $JK\mathit{\Gamma}A$ model.

\begin{figure}[t]
  \centering
  \includegraphics[width=8cm]{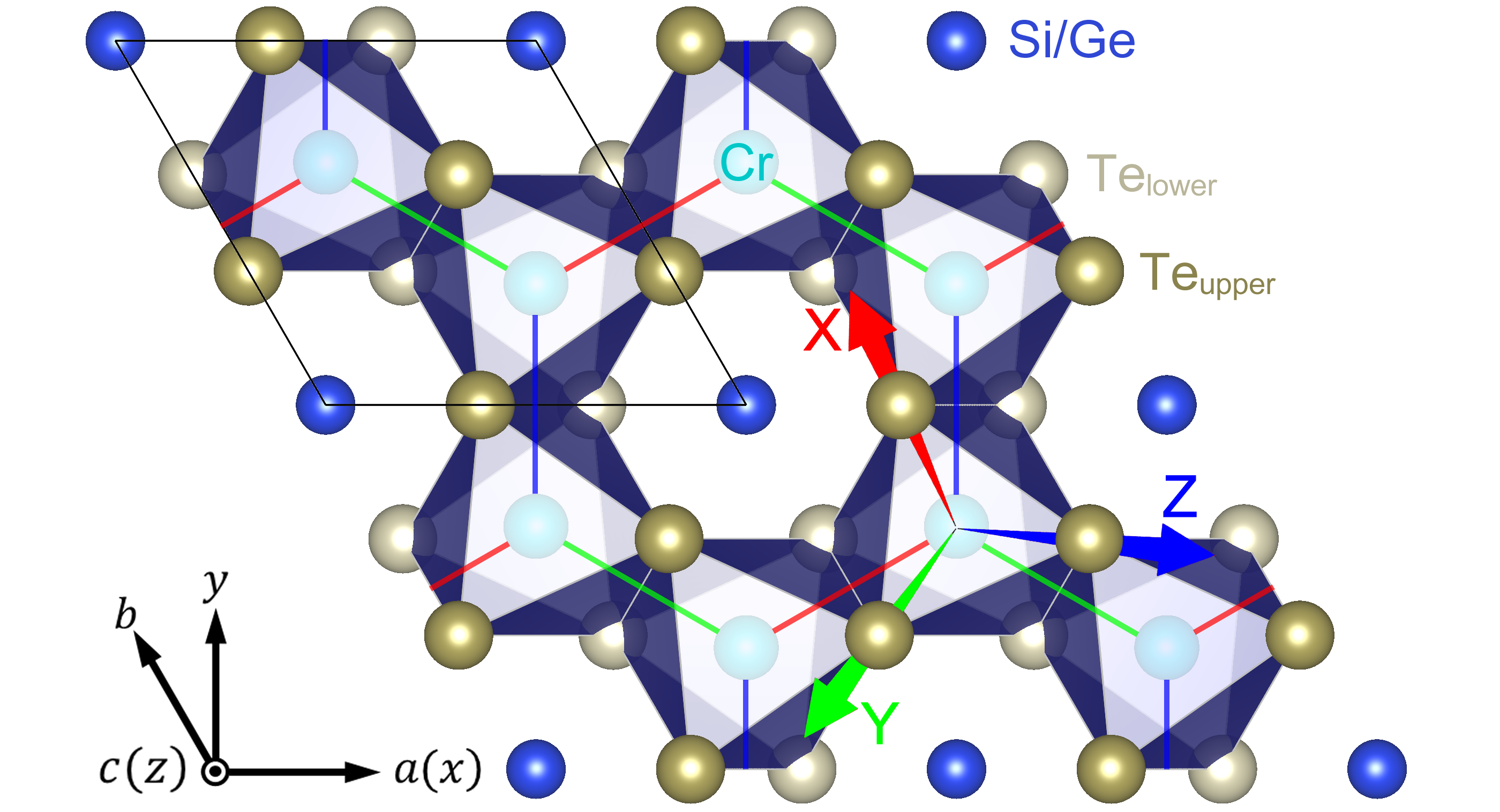}%
  \caption{ Schematization of crystal structure and the Kitaev basis of CrSiTe$_3$ monolayer. The black parallelogram marks the unit cell of the honeycomb lattice of CrSiTe$_3$ monolayer. The $\{XYZ\}$ basis of the Kitaev model is indicated by red, green and blue arrows, which is determined by L{\"o}wdin orthogonalization\cite{lowdin1951note} of the hard axes of the nearest neighbor Cr-Cr interactions. The $X$, $Y$ and $Z$ directions are found to be very close to the Cr-Te bonds \cite{note1}. 
  }
\end{figure}

Equations (1) and (2) have previously naturally reproduced and explained in Ref.\cite{xu2018interplay}  the distinct magnetic behaviors of CrI$_3$ and CrGeTe$_3$ (that is, Ising $versus$ Heisenberg behavior, respectively)\cite{samarth2017condensed,miller2017ferromagnetism}.
The approach to induce Kitaev interaction by means of heavy ligands \cite{xu2018interplay} is also evidenced by a subsequent work\cite{stavropoulos2019microscopic} that proposes achieving high-spin Kitaev physics in systems with strong SOC in anions and strong Hund's coupling in transition metal cations.
Moreover, the non-negligible effects of Kitaev interaction have also been verified by magnetization measurements on CrBr$_3$ monolayer\cite{kim2019micromagnetometry} and magnon experiments on CrI$_3$ monolayer\cite{lee2020fundamental}.
It therefore appears legitimate to explore Kitaev QSL states in Cr-based monolayers (which would then be the first Kitaev QSL candidates with partially filled 3$d$ electrons and S = 3/2), provided there is a way to make the isotropic exchange coefficient $J$  zero or nearly so while keeping $K$ finite.  The way we are going to pursue, in order to accomplish such annihilation, is to apply epitaxial strain,  since it has been shown to be an effective approach to tune the strength of exchange couplings, and thus the Curie temperature, of Cr-based systems\cite{li2014crxte}. We chose here to study CrSiTe$_3$ (CST) and CrGeTe$_3$ (CGT) under compressive strain because, as we will see, their $J$ coefficient is rather sensitive to such strain. CST has the same structure than CGT, and exhibits a similar Heisenberg behavior than CGT but with an additional slightly favoring out-of-plane anisotropy\cite{carteaux19952d,williams2015magnetic} (note that properties of CST, including the nature of its FM state, can  be well explained  by Eq (2) as well\cite{xu2018interplay}).

Technically, the elements of the exchange matrix $\mathcal{J}$, as well as the SIA coefficient $A_{zz}$, are obtained by performing density functional theory (DFT) calculations, together with the four-state energy mapping method\cite{xiang2013magnetic,xu2018interplay,xu2019magnetic}, for any investigated epitaxial strain in CST and CGT. These first-principle-derived magnetic parameters are then used as inputs of both classical Monte Carlo (MC) simulations and quantum simulations using thermal pure quantum (TPQ) states method\cite{sugiura2012thermal}.
As we will show below, such simulations reveals the existence of a strain-driven intermediate state bridging FM and antiferromagnetic (AFM) phases in CST and CGT, with this bridging state possessing many hallmarks of QSL, that are a double-peak in the specific heat-{\it versus}-temperature function and a low-temperature plateau in the temperature evolution of the entropy.
The main manuscript particularly focuses on CST, while (qualitatively-identical) results for CGT are reported in the supplemental material (SM) \cite{sm}.

Figures 2a and 2b report the behaviors of the magnetic parameters of Eq. (2) as a function of compressive strain, $\eta$, in CST,  as predicted by DFT.
Specifically, the isotropic $J$ parameter has a value of -3.42 meV at zero strain (relaxed case),  which is indicative of the FM nature of CST. But then, $J$ changes its sign at -2.41\% and therefore favors an AFM state when further increasing the magnitude of the compressive strain. It is further found that another diagonal element of the $\mathcal{J}$ matrices indicated above (i.e., $J$+$K$) changes its sign at -2.25\%.
On the other hand, the Kitaev coefficient $K$ is 0.34 meV at zero strain and only slightly decreases when increasing the magnitude of strain $\eta$.  In particular,  $K$ almost remains constant at 0.275 meV around the strain range between -2.25\% and -2.41\%.
Furthermore, when increasing the magnitude of strain $\eta$, the SIA coefficient $A_{zz}$ slightly increases and keeps the value of 0.22 meV between $\eta = -2\%$ and $-4 \%$. The opposite behaviors of $K$ and $A_{zz}$ upon varying strain, though weak, lead to the total anisotropy being in-plane when $\eta < -0.02$,
since it is previously determined that the $K$ term favors out-of-plane (through a frustration mechanism), while SIA favors in-plane for CST\cite{xu2018interplay}.
Moreover, the other terms of the $\mathcal{J}$ matrices, that are $\mathit{\Gamma}_1$ and $\mathit{\Gamma}_2$,   are found to be around an order smaller than $K$, and also change slowly with strain.
Interestingly and according to the Kitaev model\cite{kitaev2006anyons}, the vanishing of $J$ (or $J$+$K$) and the large value of $K$ hint towards the possibility of forming a QSL state near the  boundary between  FM and AFM states, as in-line with a recent study\cite{chun2015direct}.
Additionally, in our model, there is also a finite SIA term, in the strain range where $K$ is finite but $J$ (or $J$+$K$) vanishes.
The effect of such SIA term to the existence of QSL is unknown, to the best of our knowledge.

\begin{figure*}[tb]
  \centering
  \includegraphics[width=16cm]{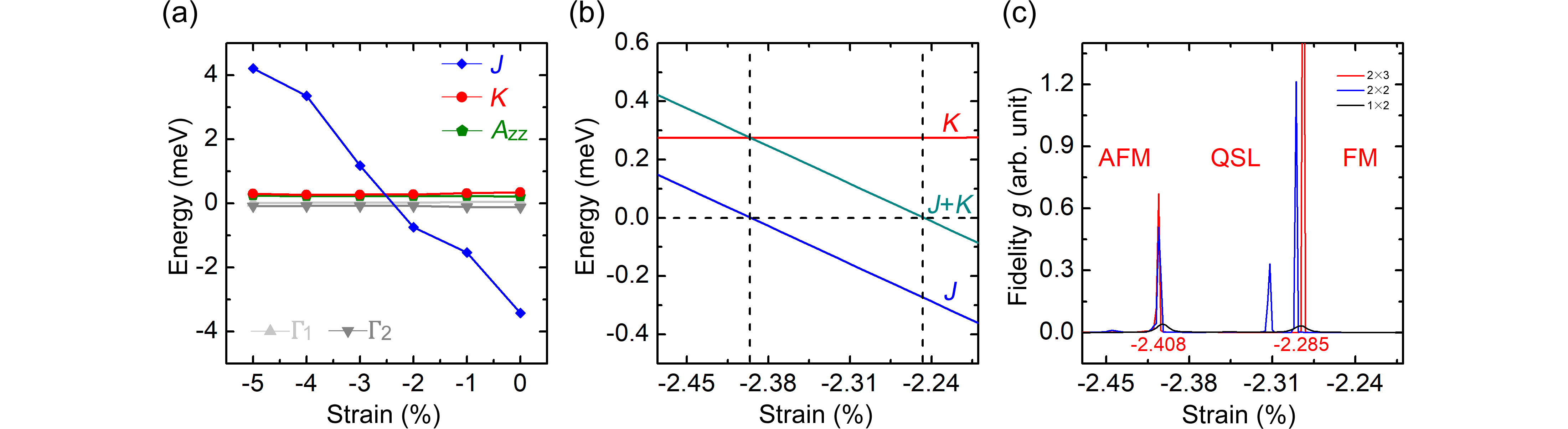}%
  \caption{Magnetic coefficients and fidelity of CrSiTe$_3$ monolayer. (a) display the evolution of magnetic coefficients as a  function of compressive strain.
  (b) further shows the evolution of $J$, $J$+$K$ and $K$ for a specific strain region. The horizontal dashed line indicates the zero in energy, while the vertical lines mark the critical strains at which $J$ or $J$+$K$ becomes zero. (c) shows the quantity of fidelity $g$ as a function of strain for different sizes of supercells.}
\end{figure*}

To determine the phase diagram of CST in its stability region and search for possible QSL states, we now compute the quantity of fidelity using the exact diagonalization (ED) method.
Practically, the DFT-extracted magnetic parameters at different strains are inserted into ED quantum simulations, from which the ground-state wave functions are obtained. The fidelity metric $g$, which measures  changes in ground-state wave functions, is then calculated as\cite{yang2012quantum},
\begin{equation}
 g = \frac{2}{{{N_s}}}\frac{{1 - F(\mu ,\delta \mu )}}{{{{(\delta \mu )}^2}}}
\end{equation}
where $F(\mu ,\delta \mu ) = \left| {\left\langle {\Psi (\mu )\left| {\Psi (\mu  + \delta \mu )} \right.} \right\rangle } \right|$ is the overlap between two ground-state wave functions at strain $\mu$ and $\mu+\delta\mu$ with $\delta\mu \to 0$; and $N_s$ is the number of sites. As $F(\mu ,\delta \mu )$ equals 0 for two states that are exactly orthogonal, a peak in fidelity $g$ will be detected at the boundary of a phase transition; on the other hand, $F(\mu ,\delta \mu )$ is nearly 1 for two states that are similar to each other and the fidelity $g$ will then show no obvious changes in such case.
The fidelity is originally a concept of quantum information, but has been recently proven to be very successful in identifying quantum phase transitions, in particular in ED simulations with limited system sizes\cite{quan2006decay,gu2010fidelity}.
It can accurately predict phase transition on the premise that the supercell (and thus $N_s$) are large enough. For small size of supercells, the scaling approach is commonly used: by increasing the size of supercells, the true peaks in fidelity $g$ become sharper, while the ``fake'' ones should gradually vanish.

\begin{figure}[bp]
  \centering
  \includegraphics[width=8cm]{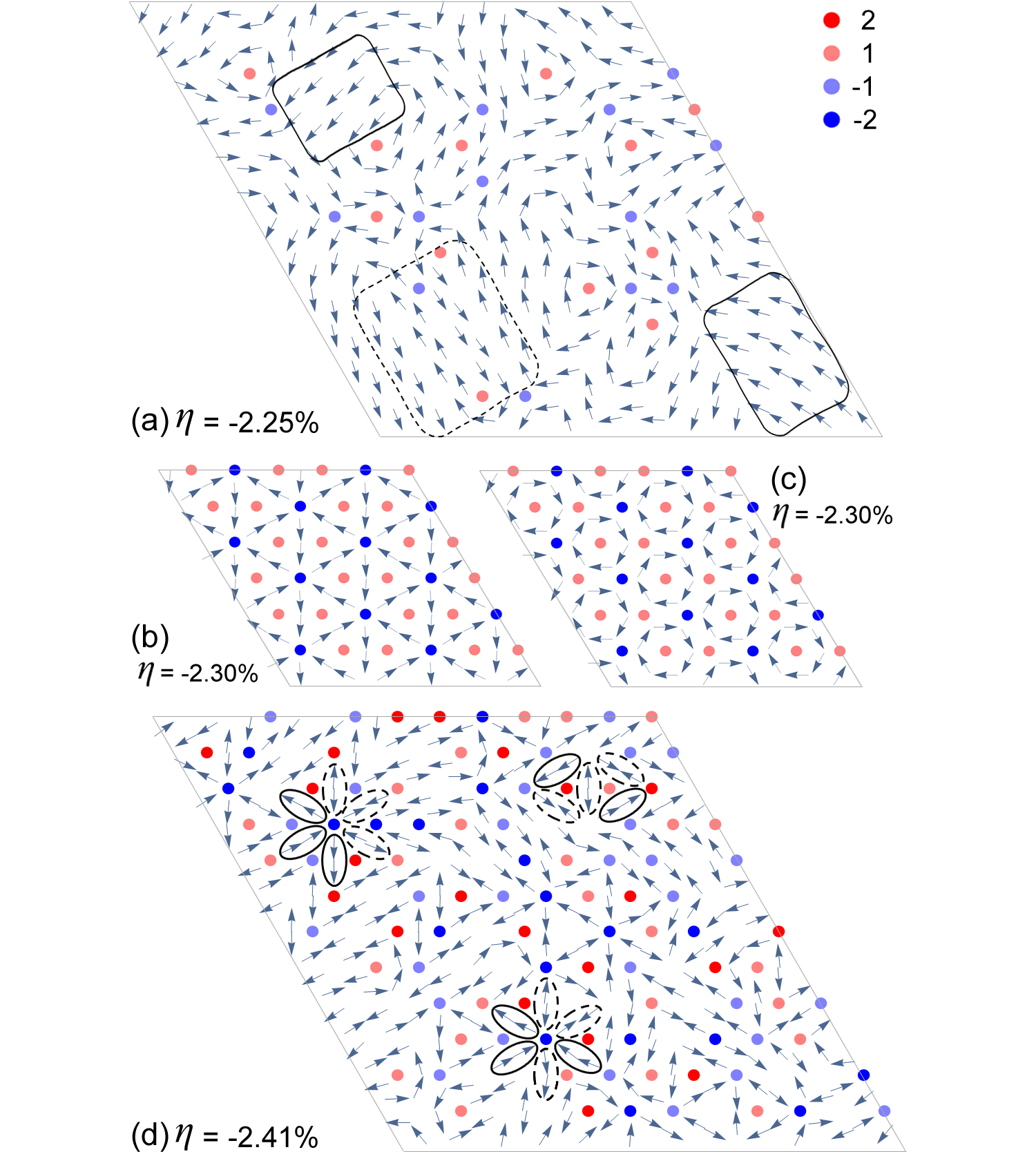}%
  \caption{Patterns of magnetic dipole moments at different strains in CrSiTe$_3$ monolayer. (a) spin patterns at $\eta$=-2.25\%, at which $J$+$K$ changes its sign. The solid and dashed rectangle marks the FM and zigzag AFM domains, respectively. Vortices and antivortices are indicated by the red and blue dots, respectively. The values represented by the dots are the vortex number, which is defined as $n=\frac{1}{2\pi}\sum_{i=1}^{6}\Delta \theta_i$, where the rotation from $i$=1 to 6 is done in an anticlockwise fashion. (b) and (c) Energetically degenerate states at $\eta$=-2.30\%. (d) spin textures at $\eta$=-2.41\%, at which $J$ changes its sign. The solid and dashed ellipses mark ferromagnetic and antiferromagnetic pairs, respectively. Note that our MC simulations are followed by a conjugate gradient algorithm, indicating that the spin patterns shown here are at global/local minima\cite{xu2020topological}.  }
\end{figure}

\begin{figure*}[tbp]
  \centering
  \includegraphics[width=16cm]{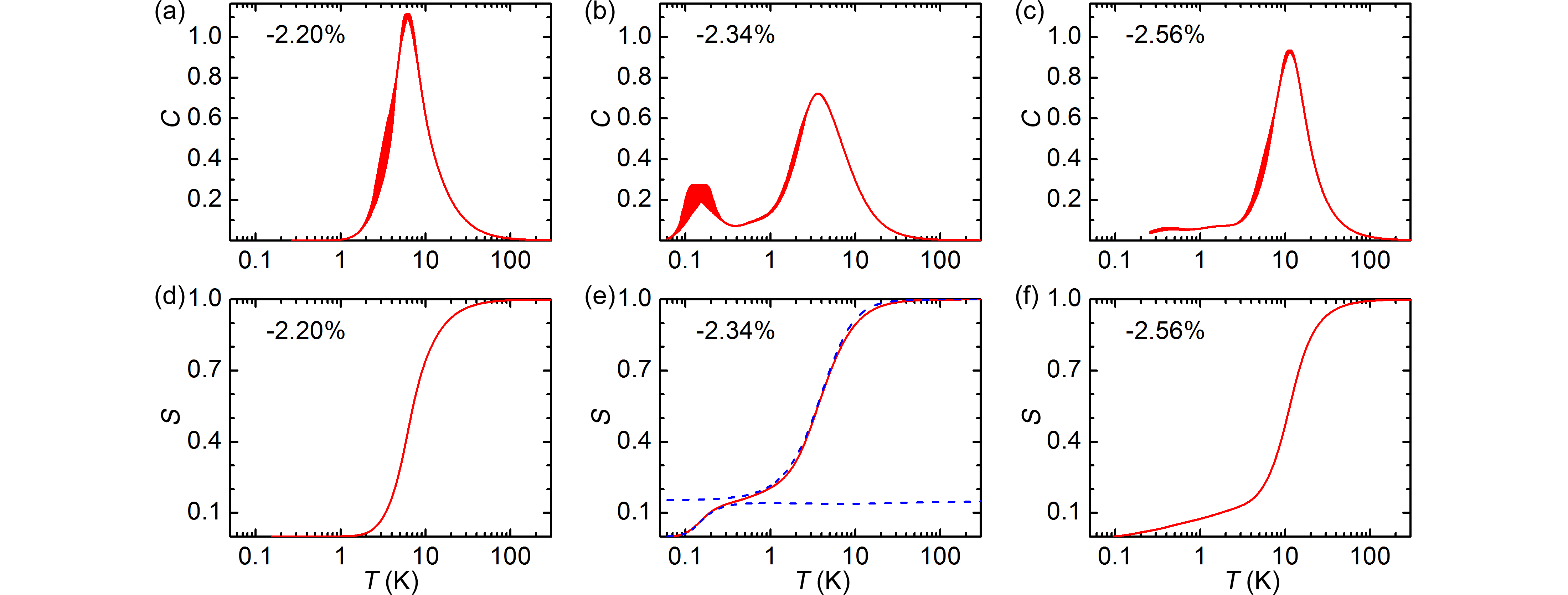}%
  \caption{Temperature evolution of specific heat and entropy of CrSiTe$_3$ monolayer at different strains. (a), (b) and (c) display the specific heat in unit of meV$\cdot$K, while (d), (e) and (f) show entropy in unit of $N$$k_B$ln4 as a function of temperature, respectively. The 2$\times$3 honeycomb lattice is used with $N$=12.}
\end{figure*}

We thus calculated the fidelity $g$ for 1$\times$2, 2$\times$2 and 2$\times$3 supercells that contains 4, 8 and 12 sites, respectively. As shown in Fig. 2(c), in the strain range extending from -2.20\% to -2.47\%, two groups of peaks are found to not only exist for all three supercells, but also become sharper with increasing size of supercells, with one group at $\eta \approx$ -2.4\% and the other group at $\eta \approx$ -2.29\%. On the other hand, another peak at $\eta \approx$ -2.31\% only exists for the 2$\times$2 supercell, and can thus be considered to be ``fake''.
It is therefore legitimate to conclude that, up to the sizes we studied, the two peaks for the 2$\times$3 supercell correspond to phase transitions. The first peak is located at -2.285\%, which is near the aforementioned -2.25\%, at which ($J$+$K$) changes its sign; and the second one is determined to be at -2.408\%, which is rather close to -2.41\% for which $J$ becomes zero.

The dipolar patterns resulting from corresponding {\it classical} MC simulations indicate that the phase at small strain is a FM state, as consistent with the observed FM state at zero strain\cite{carteaux19952d,williams2015magnetic}; while the phase at the largest compressive strain is a N\'eel-type AFM state, as also consistent with recent calculations on compressive strained CrI$_3$\cite{zheng2018tunable}. On the other hand, the  magnetic dipoles obtained from classical MC computations in the stability region of the intermediate phase (between -2.25\% and -2.41\%) exhibit a more complex pattern.
As we can see in Fig. 3a, besides the FM domain, the zigzag AFM domains begin to emerge at the phase boundary of the FM state and the intermediate phase. As further shown in Fig. 3d, both ferromagnetically coupled pairs and antiferromagnetically coupled pairs exist near the other phase boundary between the intermediate phase and AFM state. Such coexistence of FM and AFM indicates high frustration in the intermediate state.
Moreover, vortices and anti-vortices made of magnetic dipoles appear at both phase boundaries and are found to bound to each other at $\eta$=-2.25\%, which reminds us about the Berezinskii-Kosterlitz-Thouless (BKT) phase\cite{berezinsky1972destruction,kosterlitz1973ordering}. Interestingly, the BKT phase is sometimes viewed as the classical analogue of QSL\cite{yamaji2016clues,price2012critical}.
Furthermore, Figs. 3b and 3c show nearly degenerate low-energy spin patterns (among many others) within the intermediate phase.

Let us also compute two other quantities from quantum simulations that can provide signatures of QSL states, that are the specific heat $C$  and the thermal entropy $S$.
Practically, the temperature evolution of $C$ and $S$ are calculated in our systems using the TPQ method,
 \begin{equation}
C = \frac{{d\left\langle {{\phi _T}\left| {\hat H} \right|{\phi _T}} \right\rangle }}{{dT}}
\end{equation}
\begin{equation}
S = N{k_B}\ln 4 - \int_T^{ + \infty } {dT'\frac{C}{{dT'}}}
\end{equation}
where $\phi_T$ is the TPQ state at $T$ (see Ref.\cite{yamaji2016clues} for details).
According to previously established theory, spin-1/2 Kitaev QSL should exhibit two peaks in the specific heat, with the one at higher (lower, respectively) temperature being associated with localized (itinerant, respectively) Majorana fermions\cite{yamaji2016clues,baskaran2007exact,nasu2015thermal}.
Consequently, entropy release should occur twice when lowering temperature, between which a plateau should exist. For a pure higher spin Kiteav model without further terms, Majorana fermions can be maintained\cite{baskaran2008spin} and the plateau is predicted to locate at $\frac{1}{2}Nk_B$ln(2$S$+1)\cite{oitmaa2018incipient}, while the effects of $J$,  $\mathit{\Gamma}$ or SIA are yet to be determined, to the best of our knowledge.
We thus decided to to calculate the specific $C$ and thermal entropy $S$ for three specific strains, -2.20\%, -2.34\% and -2.56\%, corresponding to the different phase zones identified by the fidelity $g$.  Results are shown in Fig. 4.
For the smallest-in-magnitude strain of -2.20\%, the specific heat $C$ shows a single peak at $T$ = 6.3 K, indicating a paramagnetic (PM)-to-FM transition there. The corresponding entropy $S$ for this strain smoothly decreases to zero as the temperature decreases towards 0 K. Similarly, for the largest-in-magnitude strain of -2.56\%, a single peak in $C$ marks a PM-to-AFM transition at $T$ = 11.5 K and the temperature dependence of $S$ is rather monotonic. Strikingly, for the intermediate strain at -2.34\%, a double-peak structure is clearly identified in the specific heat $C$, with one at the higher temperature $T_h$ = 3.7 K and one at the lower temperature $T_l$ = 0.15 K. Such double-peak structure, corresponding to Majorana fermion excitations, strongly further supports the existence of Kitaev QSL around strain of -2.34\%. Moreover, as previously reported, the ratio between $T_l$ and $T_h$ can be a quantitative measure for the distance to Kitaev QSL, as $T_l/T_h$ = 0.03 for typical Kitaev QSL and $T_l/T_h$ = 0.11 for Na$_2$IrO$_3$\cite{yamaji2016clues}. Here, at strain of -2.34\%, the $T_l/T_h$ ratio is determined to be 0.04, which further emphasizes a state rather close to Kitaev QSL.
For the strain of -2.34\%, the entropy shows a clear plateau at 0.154 in unit of $N$$k_B$ln4, which is different from the 0.5 value of the pure Kitaev model\cite{oitmaa2018incipient}, but which is in line with the remarkably lowered value of 0.1935 in presence of the  $\mathit{\Gamma}_1$ term \cite{catuneanu2018path}.
As aforementioned, the distinct double-peak in $C$ and the plateau in entropy further strongly suggest that the predicted intermediate phase is indeed a Kitaev QSL.

To conclude, we have combined DFT calculations, classical MC computations and quantum simulations to  predict a possible strain-induced Kitaev QSL state in epitaxial CrSiTe$_3$ and CrGeTe$_3$ monolayers. Such  3$d$ transition metal compounds, altogether with strain engineering allowing to continuously tune the $J/K$ ratio, largely expands the scope of candidates to realize Kitaev QSL.  

\begin{acknowledgments}
We thank Dr. Ziyang Meng and Dr. Zhengxin Liu for useful discussion. This work is supported by the Arkansas Research Alliance and the Office of Basic Energy Sciences under contract ER-46612. J.F. acknowledges the support from Anhui Provincial Natural Science Foundation (1908085MA10). M.K. is supported by Priority Issue (creation of new functional devices and high-performance materials to support next-generation industries) to be tackled by using Post `K' Computer from the MEXT of Japan. Y.Q. is supported by NSFC (11874115). H.X. is supported by NSFC (11825403), Program for Professor of Special Appointment (Eastern Scholar), Qing Nian Ba Jian Program, and Fok Ying Tung Education Foundation. The Arkansas High Performance Computing Center (AHPCC) is also acknowledged.
\end{acknowledgments}


%

\end{document}